# Extreme Ultraviolet Late-Phase Flares: Before and During the *Solar Dynamics Observatory* Mission

Thomas N. Woods

LASP/University of Colorado
3665 Discovery Dr.
Boulder, CO 80303. USA
Email: tom.woods@lasp.colorado.edu

**Abstract.** The solar extreme ultraviolet (EUV) observations from the *Solar Dynamics Observatory* (SDO) have revealed interesting characteristics of warm coronal emissions, such as Fe XVI 335 Å emission, which peak soon after the hot coronal X-ray emissions peak during a flare and then sometimes peak for a second time hours after the X-ray flare peak. This flare type, with two warm coronal emission peaks but only one X-ray peak, has been named the EUV late phase (Woods *et al.*, *Astrophys. J.* **739**, 59, 2011). These flares have the distinct properties of i) having a complex magnetic field structure with two initial sets of coronal loops, with one upper set overlaying a lower set, ii) having an eruptive flare initiated in the lower set and disturbing both loop sets, iii) having the hot coronal emissions emitted only from the lower set in conjunction with the X-ray peak, and iv) having the first peak of the warm coronal emissions associated with the lower set and its second peak emitted from the upper set many minutes to hours after the first peak and without a second X-ray enhancement. The disturbance of the coronal loops by the eruption is at about the same time, but the relaxation and cooling down of the heated coronal loops during the post-flare reconnections have different time scales with the longer, upper loops being significantly delayed from the lower loops. The difference in these cooling time scales is related to the difference between the two peak times of the warm coronal emission and is also apparent in the decay profile of the X-ray emissions having two distinct decays, with the first decay slope being steeper (faster) and the delayed decay slope being smaller (slower) during the time of the warm coronal emission second peak. The frequency and relationship of the EUV late-phase decay times between the Fe XVI 335 Å two flare peaks and X-ray decay slopes are examined using three years of SDO / *EUV Variability Experiment* (EVE) data, and the X-ray dual decay character is then exploited to estimate the frequency of EUV late-phase flares during the past four solar cycles. This study indicates that the frequency of EUV late-phase flares peaks before and after each solar cycle minimum.

## Introduction

The *Extreme ultraviolet Variability Experiment* (EVE) onboard the *Solar Dynamics Observatory* (SDO: Woods *et al.*, 2012) has provided new spectral and high-cadence information about flare dynamics and energetics. Although EVE measures the full-disk irradiance in the extreme ultraviolet (EUV: 100 − 1100 Å) and the soft X-ray (SXR: 1 − 100 Å), flare events can be studied in detail as long as only one major flare is happening at a time, which happens to be most of the time. Woods *et al.* (2011) provide examples of the four major phases seen during flares with the EVE data. These phases include the impulsive phase best seen in transition region emissions such as He II 304 Å, gradual phase seen in hot coronal emissions such as the Fe XX/Fe XXIII 133 Å, coronal dimming seen in cool corona emissions such as Fe IX 171 Å and EUV late phase (ELP), which has a second, broad peak one to five hours after the main flare phases and seen best in the Fe XVI 335 Å emission. The X-ray flare classification by the *Geostationary Operational Environmental Satellite X-Ray Sensor* (GOES / XRS) is identification of the gradual phase, and the derivative of its $1-8$ Å SXR emission can be a proxy for the impulsive phase, as related to the Neupert effect (Neupert, 1968). The coronal dimming and EUV late-phase effects are only observable in the EUV emissions.

Each flare can have its own unique behavior; some flares have all four of these phases, and some flares only have the gradual phase (by definition from the X-ray flare identification by GOES /XRS). For more detailed information, Hudson (2011) reviews flare processes and phases, and Hock (2012) identifies different categories of flares based on the new SDO /EVE and *Atmospheric Imaging Assembly* (AIA) observations of hundreds of flares. Notably the eruptive flares tend to have impulsive phase, gradual phase, and coronal dimming, and some eruptive flares also have the EUV late phase. As defined by Woods *et al*. (2011), the identification of EUV late phase includes:

i) a second peak of the warm coronal emissions (Fe XV and Fe XVI) several minutes to a few hours after the GOES X-ray peak,

ii) no significant enhancements of the GOES X-ray or hot coronal emissions (*e.g.* Fe XX / Fe XXIII 133 Å) during this second peak,

iii) eruptive event as seen in the AIA images and is also seen as coronal dimming in the Fe IX 171 Å emission, and

iv) a second set of longer loops being reconnected higher than the original flaring loops and at a much later time than the first set of post-flare loops formed just minutes after the flare, as observed in AIA images.

Woods *et al*. (2011) provide an overview of about 200 flares observed during the first year of the SDO mission (May 2010 – April 2011). Of those flares, 88.5 %, 11 %, and 0.5 % of them were C, M, and X class flares, respectively. All of them had a gradual phase (as expected since flare identification starts with finding GOES X-ray peaks), 55 % had an impulsive phase, 22 % had coronal dimming, and 14 % had an EUV late-phase. In general, there is higher probability for having impulsive phase, coronal dimming (eruptive), and EUV late phase for the larger flares.

From examining hundreds of flares during the SDO mission, it has been noticed that there is an interesting dual-decay behavior of the GOES X-ray time series during the flare decline if the flare was identified to have an EUV late phase. Cargill *et al*. (1995) and others have shown that coronal-loop cooling time is proportional to loop length. More specifically for EUV late-phase flares, Liu *et al*. (2013) provide detailed analysis and results for the cooling times and loop lengths for two late-phase flare events for both the main-phase loops and the late-phase loops. With the X-ray decay rate being related to post-flare loops cooling down, then an X-ray flare with two distinct decay periods may indicate two different sets of post-flare loops. There is an expectation for EUV late-phase flares that the first steep slope, right after the X-ray peak, is related to the cooling time of the main-flare loops and that the second, less-steep slope is related to the slower cooling time of the longer, overlying loops associated with an EUV late-phase flare. Some example flares are shown in Figure 1 with the second decay slope highlighted as potential indicator for an EUV late-phase. There is also the possibility that long decay events (LDEs) and extended-flare-phase phenomena (*e.g.* Hudson, 2011) are contributing to extended heating in the main flare loops and causing dual-slope behavior in the X-ray time series but without requiring a second set of loops as needed for the EUV late phase. Thus all dual-decay X-ray flares may not be EUV late-phase flares; therefore, comparisons of X-ray flares to the SDO EUV measurements are important to understand the possible relationship of dual-decay X-ray flares to EUV late-phase flares.

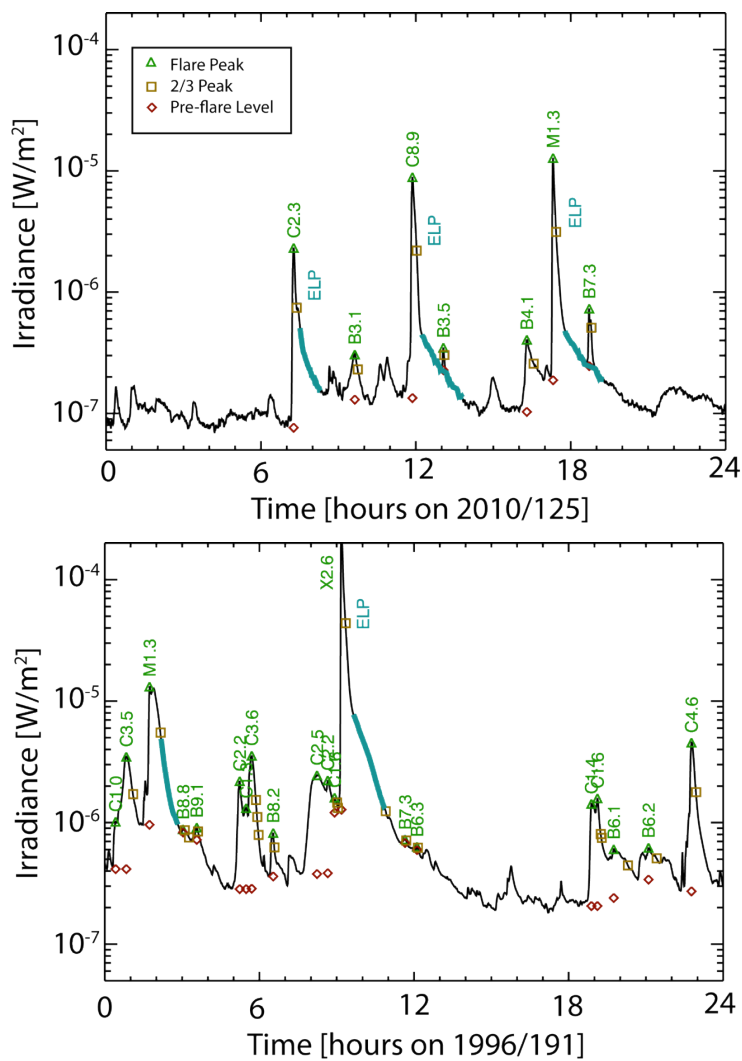

**Figure 1.** Examples of Identifying Possible EUV Late-Phase (ELP) Flares with the GOES X-ray Data. The dual-decay algorithm identifies and labels the second, slower slope in the GOES X-ray data with blue symbols, and these are also labeled as potential ELP events. The top panel shows a day during the SDO mission (4 May 2010), and those identified as ELP flares are consistent with ELP flares identified by Woods *et al.* (2011). The bottom panel shows a day before the SDO mission (8 July 1996). For each flare considered in the algorithm, the flare peak is labeled in green text, the pre-flare level is labeled as a red diamond, and the main-flare end time is labeled as an orange square.

This dual-decay behavior of some GOES X-ray flares is examined over the interval 1974 – 2013 to study if EUV late-phase flare frequency of occurrence might change over the solar cycle. The SDO/EVE concurrent measurements provide validation for how accurate this dual-decay behavior of the X-ray flares is for identifying EUV late-phase flares. As shown later, this dual-decay behavior of the X-rays is not highly accurate for identifying EUV late-phase flares when the Sun is very active, but it does appear to be a reasonable proxy for EUV late-phase flares prior to having SDO observations. The GOES X-ray record is then examined to study the frequency of occurrence of dual-decay X-ray flares over the past four solar cycles.

A motivation for this study of the EUV late-phase flare frequency over the solar cycle is to improve upon the understanding on how the solar EUV radiation can influence Earth's ionosphere variations. It is well known that the solar EUV radiation is a major energy source to create and drive the ionosphere; for example, the EUV radiation at wavelengths shorter than 400 Å influence the ionosphere F layer peak density and height (*e.g.* Liu *et al.*, 2011). In addition, geomagnetic activity and ion-neutral dynamics also influence the ionosphere variations. Some ionospheric studies on solar EUV variability only use proxies such as the 10.7-cm radio flux ($F_{10.7}$) for daily variations and GOES X-ray irradiances for flare variations (*e.g.* Liu *et al.*, 2011; Xiong *et al.*, 2013), but such an approach can not account for the EUV late-phase energy input that can arrive one to five hours later than the main-flare peak as detected in the X-rays. If the EUV late-phase flare frequency does have a solar-cycle dependence, then such a result can help focus the time periods to study the effects of the EUV late phase on the ionosphere.

**Algorithm for Finding Dual Decay Flares**

The algorithm to find X-ray flares that have dual decays starts by first finding all flare peaks and also identifying the pre-flare level and the main-flare end time for each flare. The flare end time is defined here as the time when the logarithm of the X-ray irradiance [log(GOES)] is at the two-thirds level between the flare peak and pre-flare level. A limit of six hours is used for the main-flare end time in case there are numerous flares in a row where the X-ray background level does not return to the pre-flare level.

With all of the X-ray flares identified in this way, then a search for dual decays is performed. As the second decay is at much lower irradiance levels than the flare peak, one check is that the main-flare peak for X-ray irradiance has to be greater than a C1 and at least four times higher than the X-ray background. As the EUV late phase is defined not to be contaminated with a new flare event, another check is that there can not be any flares larger than a quarter (or 25 %) of the flare peak for three hours after the initial flare peak. Based on EUV late phases having a delayed peak by one to five hours, a third check is that the second decay period has to be prior to returning to the X-ray pre-flare level and at least 60 minutes after the flare peak. The start of the second decay period is after the end of the main-flare period when there is a change in the decay slope, and the end of the second decay period is constrained by returning to the pre-flare level, the start of another flare, or six hours as an upper limit. The final check on the existence of the second decay period is that the second decay slope is smaller in magnitude than the first decay slope and that log(GOES) slope is more negative than -0.005 per minute. An exponential decay can simply be a linear fit if the logarithmic function is used, so the slopes for both decay periods are fit on log(GOES) after a five-minute smoothing is applied.

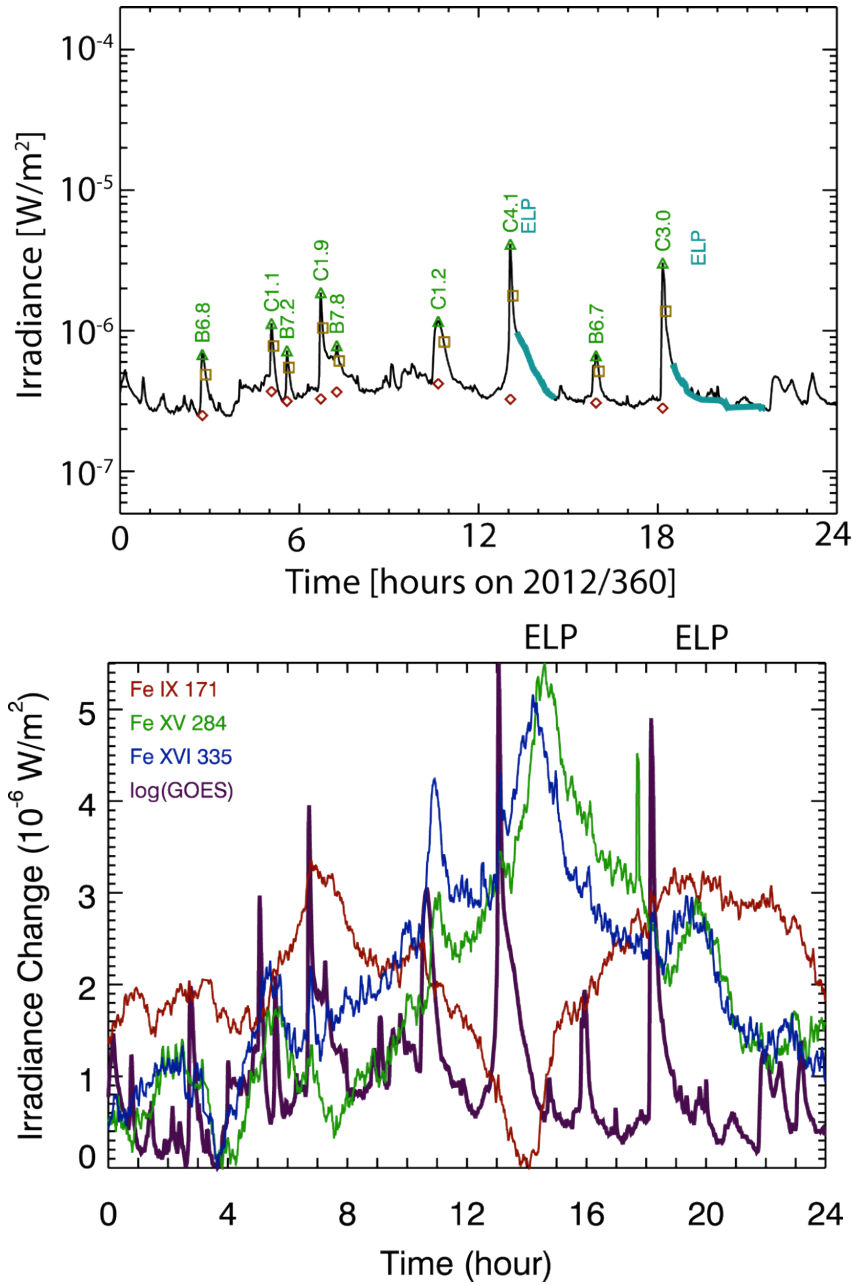

**Figure 2.** Validation of the Dual-Decay Algorithm Result with EVE Data. The results from this algorithm are possible during the SDO mission using the behavior of the Fe XV 284 Å and Fe XVI 335 Å emissions as indicators for the EUV Late Phase (ELP) and the Fe IX 171 Å emission as indicator for an eruptive event. The top panel shows the dual-decay algorithm results (as shown in Figure 1 for other days). The bottom panel shows the irradiance changes for the EVE data and the log of the GOES X-ray. The C4.1 and C3.0 flares are identified as possible ELP flares by the dual-decay algorithm. Both are confirmed to be ELP flares with the EVE data because the Fe IX 171 Å has a small dip (coronal dimming) during the peak of the GOES X-ray, and both the Fe XV 284 Å and Fe XVI 335 Å emissions have a first, small peak near the time of the GOES X-ray peak and then a second, broad peak after the X-ray peak.

A summary of the dual-decay flare search is given in Table 1, and some example time series of the search are shown in Figures 1 and 2. The next section describes how well the second decay period in the X-ray time series corresponds to identified EUV late-phase flares in the SDO/EVE data.

**Table 1.** Summary of Results for Dual-Decay X-ray Flares (DDXF)

| Flare Class | Total Flares GOES 1974 − 2013 | Frequency DDXF GOES 1974 − 2013 | Total Flares SDO Mission 2010 − 2013 | Frequency DDXF SDO Mission 2010 − 2013 |
|---|---|---|---|---|
| C | 65 328 | 5.4 % | 5163 | 8.4 % |
| M | 9665 | 22 % | 318 | 41 % |
| X | 869 | 39 % | 19 | 74 % |
| Total | 75 862 | 7.9 % | 5500 | 10.5 % |

**Validation during SDO Mission**

There are hundreds of flares during the SDO mission to examine manually, so a simplified version of the EUV late-phase definition is considered here using only EVE and GOES data. The ELP rule #1 (second warm-corona peak) is verified with the Fe XV 284 Å and Fe XVI 335 Å emissions having two peaks, one peak near the main-flare peak and a broader peak one to six hours later. The ELP rule #2 (no significant enhancements of the X-ray during second peak) is forced through the selection of the second decay period in the X-ray data that requires that there are no significant flares for three hours after the main-flare peak. The ELP rule #3 (eruptive event) is verified if the Fe IX 171 Å has coronal dimming. The ELP rule #4 (two sets of coronal loops) is not verified directly with solar images for this analysis but is assumed to exist if dual-decay periods were found in the X-ray time series after the flare peak. The ELP rule #4 was verified with AIA images for the flares during the period of 30 April 2010 to 7 March 2011 as reported by Woods *et al.* (2011).

An example of this validation is shown in Figure 2 for day 24 December 2012. For the C4.1 flare near 13 UT, the Fe IX 171 Å has a large dip (coronal dimming) that starts before the GOES X-ray flare peak, and both the Fe XV 284 Å and Fe XVI 335 Å emissions have a small peak near the time of the GOES X-ray peak and then a second, broad peak after 14:00 UT. This is also similar for the C3.0 flare near 18:00 UT but with the Fe IX 171 Å dip being smaller and not lasting as long as for the C4.1 flare. Both of these flares are considered to be confirmed as ELP flares.

There are three ELP identification categories listed in Table 2 for the dual-decay X-ray flares. The dual-decay X-ray flare is considered to be confirmed as an ELP flare if the Fe IX 171 Å emission has coronal dimming (eruptive event) and if the Fe XV 284 Å and Fe XVI 335 Å emissions both have two peaks. The dual-decay X-ray flare is not considered to be an ELP flare if the Fe IX 171 Å emission does not have coronal dimming or if neither the Fe XV 284 Å emission nor the Fe XVI 335 Å emission has a second peak. It is considered to be a potential

ELP event if the Fe IX 171 Å emission has coronal dimming and if only one of the Fe XV 284 Å or Fe XVI 335 Å emissions has a second peak.

The first validation period listed in Table 2, being from 30 April 2010 through 7 March 2011, shows very good success in confirming ELP flares for the identified dual-decay X-ray flares. The second validation period, being from 8 March 2011 through 29 April 2013, is not as accurate in identifying ELP flares from the dual-decay X-ray flares. It is clear that the second period is during a period of higher solar activity with many more flares than the first period. From manually examining all of these time series, it became obvious that the technique for identifying dual-decay X-ray flares can be corrupted with additional flares appearing one to six hours after the initial flare. With the second decay level often being a factor of ten less bright than the initial flare peak, even small B-class flares can affect the algorithm searching for the second decay period after each flare. Combining the results from both time periods, it appears that this technique for finding X-ray flares with dual slopes has only 50 % accuracy for identifying ELP flares. Because of this poor accuracy, the SDO observations are definitely the choice for identifying ELP flares. However, this dual-decay X-ray algorithm is considered good enough to examine whether there are any obvious patterns in ELP flare frequency over the solar cycle, as discussed in the next section.

**Table 2.** Comparison of Dual-Decay X-ray Flares to EUV Late-Phase (ELP) Flares. The three numbers in square brackets [ ] are the number of C, M, and X class flares for each category.

| ELP Flares | 30 April 2010 – 7 March 2011 | 8 March 2011 – 29 April 2013 |
|---|---|---|
| Positive Identification | 41 (57 %) | 101 (24 %) |
| [C, M, X class flares] | [28, 5, 0] | [66, 30, 5] |
| Potential Identification | 5 (7 %) | 75 (18 %) |
| [C, M, X class flares] | [5, 0, 0] | [58, 16, 1] |
| False Detection | 26 (36 %) | 238 (57 %) |
| [C, M, X class flares] | [25, 1, 0] | [193, 41, 4] |
| Total Number | 72 | 414 |

**Results before SDO Mission**

This dual-decay X-ray algorithm is applied for the GOES X-ray record from 1974 to 2013, and those results are compiled into annual summaries as classified by C, M, and X flares and the total of those flares. The years 1974 and 2013 do not include a full 12 months; thus the end-point results may be even less accurate. Figure 3 includes three plots of the trends from this analysis.

*Figure 3 top and middle panels:*

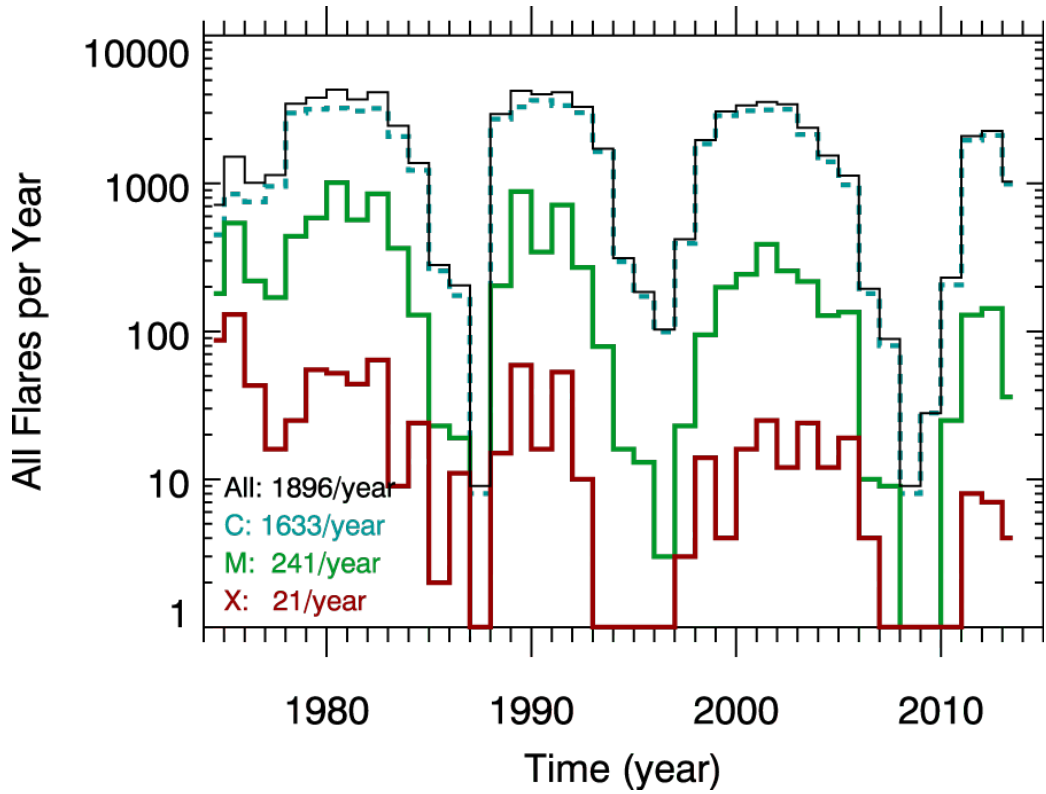

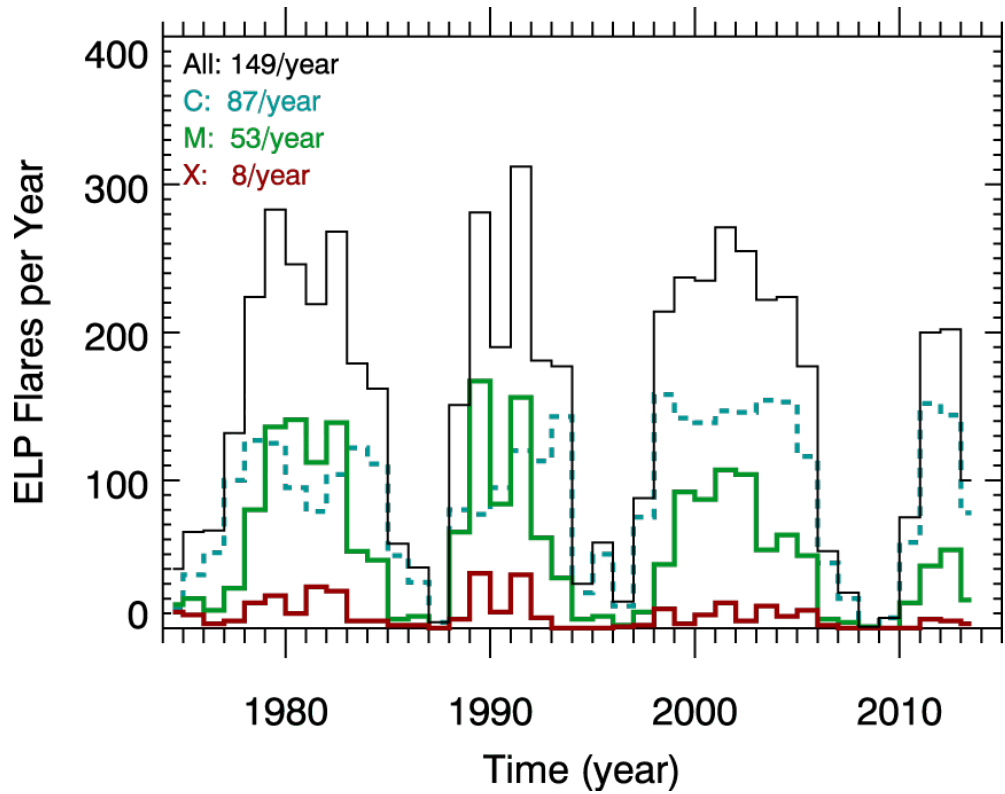

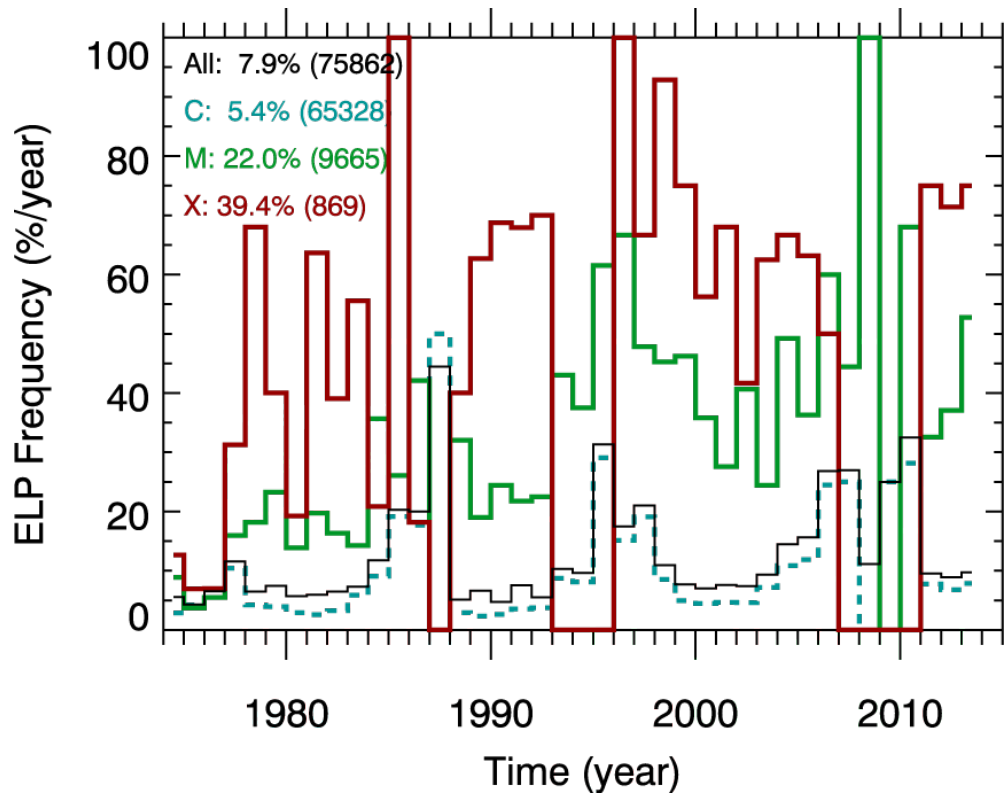

**Figure 3.** Annual Variations of Potential EUV Late-Phase (ELP) Flares. The dual-decay algorithm results are compiled into annual averages for C-, M-, and X-class flares separately. The top panel shows the number of all flares identified per year in the GOES X-ray data. The middle panel shows the number of potential ELP flares identified per year. The solar-cycle trend is clear in both the top and middle panels, and there is an interesting trend that there are fewer flares in sequential cycle maxima since the 1990s. The bottom panel shows the ratio of the number of ELP flares to all flares, and this ratio represents the frequency that flares might be ELP flares. The ELP frequency for all flares and C-class flares has peaks right before and right after solar-cycle minimum and is lower during solar-cycle maximum. The ELP frequency for M- and X-class flares can be more confusing because there are very few, or no, large flares for years during and nearby solar-cycle minimum.

The total number of flares found in the GOES X-ray record per year is shown in the top panel of Figure 3. This result is from searching the GOES record for flare peaks and thus is regardless if a dual-decay X-ray flare is found or not. This plot clearly shows that more flares appear during solar-cycle maximum than during minimum, as expected; however, the long-term trend from maximum to maximum is interesting in showing a decrease in the number of flares since 1990. That is, the number of flares during Solar Cycle 23 maximum in 2000 − 2003 is lower than the flares during Cycle 22 maximum in 1989 − 1992. This decrease is seen for all levels (C, M, and X) and the decrease for X-class flares is larger than the M-class flares, which in turn has a decrease larger than the C-class flares. Furthermore, the number of flares during Cycle 24 maximum in 2011 − 2013 is even more reduced as compared to Cycle 23 maximum. Of course, the Solar Cycle 24 maximum period is not necessarily over yet, but many of the solar proxies, such as sunspot number and 10.7-cm radio flux ($F_{10.7}$), appear to have already had their peak for this cycle. Some cycles have two peaks during maximum, often separated by a couple years, so Cycle 24 maximum may not have been observed yet. This downward trend for flares during cycle maximum is also seen in the cycle-minimum trend, although the minimum trend is less clear with the number of M and X flares sometimes being near zero near cycle minimum. The wider than usual cycle minimum in 2007 − 2010 may also contribute to the trend for the flares during cycle minimum.

One concern with any long-term solar trend is if the performance and calibration of the instruments could be changing enough to influence the observed trend. The series of GOES/XRS instruments have been cross-calibrated with concurrent measurements from multiple GOES satellites, and their calibrations are thought to have an accuracy of at least 30 % (private communication, R. Viereck and W. Neupert, 2013). Furthermore, we have flown calibration rockets for *Thermosphere Ionosphere Mesosphere Energetics and Dynamics/Solar EUV Experiment* (TIMED/SEE) and SDO/EVE for the past ten years, and these comparisons to GOES/XRS irradiances have consistently shown differences of less than 30 %. The downward trend in number of flares is almost a factor of two for M and X flares from one maximum to the next, so instrumental effects at the 30 % level cannot be a major concern for these trend results. Then the logical conclusion is that the downward trend is due to a long-term decrease in solar activity. Russell *et al*. (2010) and Fröhlich (2011) have reported that solar activity has been decreasing since the 1990s, and these results also support this conclusion.

The results for the dual-decay X-ray flares are shown in the middle and bottom panels of Figure 3. The middle panel shows the number per year, and bottom panel shows the relative number (percent) relative to the total number of flares per year. These results are considered to be a proxy for ELP flares, but with the understanding that there is about 50 % uncertainty in these values and also larger uncertainty during cycle maximum. The middle panel of number of potential ELP flares follows a similar pattern as the total number of flares in having more events during cycle maximum than during minimum. The bottom panel may be more enlightening because it represents the frequency that flares might be ELP flares. The ELP frequency for all flares and C-class flares have peaks before and after solar cycle minimum of about 20 % to 30 % and is lower during solar cycle maximum at about 10 %. The cycle minima were in 1985, 1995, and 2008. Even with the 50 % uncertainty for these results, this result of solar-cycle dependence for ELP flare frequency is at least a three-σ result. Furthermore, this behavior is consistently seen for all four solar cycles.

The ELP flare frequency for M and X flares can be more confusing because there are very few, or no, large flares for years during cycle minimum. Ignoring the periods near cycle minima when there are often no X-class flares, the frequency for ELP flares for larger flares is about 50% and is significantly higher than the results for the C-class flares. Many X-class flares are two-ribbon flares or long-duration events (LDEs) and are associated with larger and more complex active regions. The confirmed ELP flares during the SDO mission are associated with the more complex active regions that have multiple sets of coronal loops, so perhaps it is reasonable that M- and X-class flares could have more ELP flares than the C-class flares which often associate with more simple active-region configurations.

**Conclusions**

While not the initial intention for this study, one important result from this study is that the number of flares has been decreasing since the 1990s. The downward trend is notably greater for the larger M- and X-class flares, being almost a factor of two decrease per cycle. The solar-irradiance variability for Cycle 24 is almost half of the Cycle 23 variability, so it is reasonable that the number of flares during Cycle 24 is reduced relative to Cycle 23. What is somewhat of a surprise is that there is also a notable decrease of flares in Cycle 23 relative to the number of flares in Cycle 22, especially when considering that the solar-irradiance variability in Cycles 22 and 23 is comparable.

This study's primary goal was to see if the dual-decay behavior of some X-ray flares could be a proxy for ELP flares. Through validation with SDO/EVE data, the success of dual-decay X-ray flares being confirmed as ELP flares is only about 50 %. The dual-decay X-ray flares could also be the result of extended heating in the main-flare loops without requiring a second set of loops as needed for the EUV late phase; thus all dual-decay X-ray flares do not have to be EUV late-phase flares. A possible conclusion could be that the dual-decay behavior is just a poor proxy for ELP flares; however, the manual examination of hundreds of flares with GOES and SDO/EVE data indicates that cases not contaminated with other flares within six hours of the main flare are much more accurate in confirming ELP flares when a dual-decay X-ray flare is detected. Some of the analysis inaccuracy is due to the algorithm itself, with fixed values for flare and background intensities, and also due to complications of having multiple flares near each other in time. With the understanding that these results have only 50 % accuracy, it is assumed for the rest of the discussion that the dual-decay X-ray flares are reasonable proxies for ELP flare events. The analysis then examined all of the GOES X-ray data from 1974 to 2013 to find cases of dual-decay X-ray flares.

This study indicates that the frequency of EUV late-phase flares peaks before and after each solar-cycle minimum and has a minimum frequency of occurrence during cycle maximum. This behavior is consistently seen over four cycles. Many ELP flares were seen in the early part of the SDO mission during the initial rise of Cycle 24, and not as many ELP flares were observed during Cycle 24 maximum. This study suggests that this behavior of ELP flares during the SDO mission is normal behavior over the solar cycle. In addition, this frequency peak after minimum also helps to explain why the summary results in Table 1 have higher frequency during the SDO mission relative to the four solar cycles because the three years of the SDO mission have occurred primarily during a period right after the Cycle 23/24 minimum.

Why does the frequency of dual-decay X-ray flares, as a proxy for ELP flares, peak before and after cycle minimum? This initial study without a detailed look at solar images cannot properly answer this question, nor can we speculate on how active region complexity evolves over the solar cycle to allow for peaks in the frequency of ELP flares. While there are not enough observations of the solar corona to study the configuration of the active regions for all (most) flare events back to 1974, the SDO / AIA, SDO / *Helioseismic and Magnetic Imager* (HMI), *Solar and Heliospheric Observatory* (SOHO) / *Extreme ultraviolet Imaging Telescope* (EIT), and *SOHO* / *Michelson Doppler Imager* (MDI) solar images may be adequate to study the EUV coronal structures and magnetic configurations of the active regions for most flare events back to 1996, thus covering almost two solar cycles.

**Acknowledgements**

This research is supported by NASA contract NAS5-02140 to the University of Colorado. I thank the SDO/EVE and GOES/XRS teams for providing the high-quality data used in this study.

**References**

Cargill, P.J., Mariska, J.T., Antiochos, S.K.: 1995, Cooling of solar flare plasmas. I. Theoretical considerations, *Astrophys. J.* **439**, 1034.

Fröhlich, C.: 2011, A four-component proxy model for total solar irradiance calibrated during solar cycles 21-23, *Contrib. Astron. Obs. Skalnaté Pleso* **35**, 1.

Hock, R.A.: 2012, *The Role of Solar Flares in the Variability of the Extreme Ultraviolet Solar Spectral Irradiance*, PhD Thesis, University of Colorado, Boulder.

Hudson, H.S.: 2011, Global properties of solar flares, *Space Sci. Rev.* **158**, 5, doi: 10.1007/s11214-010-9721-4.

Liu, K., Zhang, J., Wang, Y., Cheng, X.: 2013, On the origin of the extreme-ultraviolet late phase of solar flares, *Astrophys. J.* **768**, 150.

Liu, L., Wan, W., Chen, Y., and Le, H.: 2011, Solar activity effects of the ionosphere: a brief review, *Chinese Sci. Bull.* **56**, 1202.

Neupert, W.M.: 1968, Comparison of solar X-ray line emission with microwave emission during flares, *Astrophys. J. Lett.* **153**, L59.

Russell, C.T., Luhmann, J.G., Jian, L.K.: 2010, How unprecedented a solar minimum?, *Rev. Geophys.* **48**, RG2004, doi:10.1029/2009RG000316.

Woods, T.N., Hock, R., Eparvier, F., Jones, A.R., Chamberlin, P.C., Klimchuk, J.A., Didkovsky, L., Judge, D., Mariska, J., Warren, H., Schrijver, C.J., Webb, D.F., Bailey, S., Tobiska, W.K.: 2011, New solar extreme ultraviolet irradiance observations during flares, *Astrophys. J.* **739**, 59, doi:10.1088/0004-637X/739/2/59.

Woods, T.N., Eparvier, F.G., Hock, R., Jones, A.R., Woodraska, D., Judge, D., Didkovsky, L., Lean, J., Mariska, J., Warren, H., McMullin, D., Chamberlin, P., Berthiaume, G., Bailey, S., Fuller-Rowell, T., Sojka, J., Tobiska, W.K., and Viereck, R.: 2012, The *EUV Variability Experiment* (EVE) on the *Solar Dynamics Observatory* (SDO): Overview of science objectives, instrument design, data products, and model developments, *Solar Phys.* **275**, 115, doi: 10.1007/s11207-009-9487-6.

Xiong, B., Wan, W., Ning, B., Ding, F., Hu, L., Yu, Y.: 2013, A statistic study of ionospheric solar flare activity indicator, *Space Weather*, doi: 10.1002/2013SW001000.